\documentclass[journal]{IEEEtran}

\IEEEoverridecommandlockouts

\usepackage{graphics} 
\usepackage{epsfig} 
\usepackage{times} 
\usepackage{amsmath} 
\usepackage{amssymb}  
\usepackage{multirow}
\usepackage[ruled,linesnumbered]{algorithm2e}
\usepackage{epstopdf} 
\usepackage{subfigure} 
\usepackage{stfloats} 
\usepackage{bm}
\usepackage{color}
\usepackage{amsthm}
\usepackage{cite}
\usepackage{array}
\usepackage{float}

\newtheorem{definition}{\indent Definition}
\newtheorem{remark}{Remark}
\begin{document}
	
\title{Curriculum-based Deep Reinforcement Learning for Quantum Control}
\author{Hailan~Ma, Daoyi~Dong, Steven X. Ding, Chunlin~Chen
\thanks{This work was supported by the National Natural Science Foundation of
		China (No.62073160 and No.61828303), the Australian Research
		Council's Discovery Projects funding scheme under Project DP190101566, the Alexander von Humboldt Foundation of Germany, and  the U. S. Office of Naval Research Global under Grant N62909-19-1-2129.}
\thanks{H. Ma is with the School of Engineering and Information Technology,
		University of New South Wales, Canberra, ACT 2600, Australia (email: hailanma0413@gmail.com).}	
\thanks{D. Dong is with the School of Engineering and Information Technology,
	University of New South Wales, Canberra, ACT 2600, Australia, and with Institute for Automatic Control and Complex Systems (AKS), University of Duisburg-Essen, 47057 Duisburg, Germany (email: daoyidong@gmail.com).}	
\thanks{S. X. Ding is with Institute for Automatic Control and Complex
Systems (AKS), University of Duisburg-Essen, 47057 Duisburg, Germany
(email: steven.ding@uni-due.de).}
\thanks{C. Chen is with the Department of Control and Systems Engineering, School of Management and Engineering, Nanjing University, Nanjing 210093, China (e-mail: clchen@nju.edu.cn).}
}
\maketitle
\begin{abstract}
Deep reinforcement learning has been recognized as an efficient technique to design optimal strategies for different complex systems without prior knowledge of the control landscape. To achieve a fast and precise control for quantum systems, we propose a novel deep reinforcement learning approach by constructing a curriculum consisting of a set of intermediate tasks defined by a fidelity threshold. Tasks among a curriculum can be statically determined using empirical knowledge or adaptively generated with the learning process. By transferring knowledge between two successive tasks and sequencing tasks according to their difficulties, the proposed curriculum-based deep reinforcement learning (CDRL) method enables the agent to focus on easy tasks in the early stage, then move onto difficult tasks, and eventually approaches the final task. Numerical simulations on closed quantum systems and open quantum systems demonstrate that the proposed method exhibits improved control performance for quantum systems and also provides an efficient way to identify optimal strategies with fewer control pulses.
\end{abstract}


\section{Introduction}\label{sec:introduction}
Manipulating quantum systems with high efficiency  \cite{nielsen2010quantum,wiseman2009quantum,dong2010quantum} is a  major challenge in developing quantum technology and provides recipes for 
many applications such as quantum computation \cite{nielsen2010quantum},  quantum communication, and quantum sensing \cite{giovannetti2011advances}. 
To achieve quantum operations with high efficiency, control methods, such as optimal control theory \cite{dong2010quantum}, closed-loop learning control algorithms \cite{rabitz2000whither} and Lyapunov control approaches \cite{kuang2017rapid}, have been developed for manipulating quantum systems. Among them, gradient algorithms have been used for numerically finding an optimal field \cite{khaneja2005optimal}. Evolutionary computing methods such as genetic algorithm (GA) and differential evolution (DE) have been utilized in  optimizing molecular systems \cite{ma2015differential, ma2017quantum,dong2019differential,dong2015sampling}. However, in many practical applications, the gradient information may not be easy to obtain and evolutionary algorithms usually involve a process of evolving a population and tend to be time-consuming when solving complex problems. 

Machine learning has attracted increasing attention in recent years owing to its powerful computing capability and has been gradually applied in various quantum tasks in recent years \cite{biamonte2017quantum}.
 Particularly, reinforcement learning (RL) \cite{sutton2018reinforcement} offers a considerable advantage in controlling realistic systems without constructing a reliable effective model. For example, a fidelity-based probabilistic Q-learning approach that incorporates fidelity into updating Q-values and action selection probabilities achieves a better balance between exploitation and exploration when dealing with quantum systems \cite{chen2013fidelity}. It has also been found that RL-aided approaches succeed in identifying variational protocols with nearly optimal fidelity, even in the glassy phase, where optimal state manipulation is exponentially hard \cite{bukov2018reinforcement}. Recently, the combination of RL and deep learning \cite{lecun2015deep,goodfellow2016deep}, i.e., deep reinforcement learning (DRL), exhibits effective representations for learning agents \cite{mnih2015human,li2019deep,zhao2018special,wang2020coordinated} and therefore achieves efficient control of different quantum systems \cite{chen2013fidelity,bukov2018reinforcement,
zhang2019does,mackeprang2019reinforcement,an2019deep,niu2019universal,fosel2018reinforcement,xu2019generalizable,bharti2019teach,chen2019extreme,porotti2019coherent,haug2020classifying}. For example, a network-based ``agent” is designed for discovering complete quantum-error-correction strategies to protect a collection of qubits against noise \cite{fosel2018reinforcement}. With the help of DRL,  nearly extreme spin squeezing with a one-axis twisting interaction is achieved using merely a handful of rotation pulses \cite{chen2019extreme}. In addition, DRL realizes efficient and precise gate control \cite{an2019deep} and exhibits strong robustness when dealing with parameter fluctuations \cite{niu2019universal}.  Besides, DRL methods are also employed for discovering quantum configurations of measurement settings and quantum states for violating various Bell inequalitie\cite{bharti2019teach}, and designing generalizable control for quantum parameter estimation \cite{xu2019generalizable,schuff2020improving}.


When optimizing the control fields using RL, the learning process is usually quite slow due to complex dynamics, or sparse reward signals.
In that case, the early transitions are likely to terminate on states that are easy to reach, and those states that are difficult to reach are usually found later in the training process \cite{narvekar2020curriculum}. However, in practical settings, these easy-to-reach states may not provide a reward signal, which might hinder the training process of the DRL agent. In addition, it usually takes millions of episodes for an RL agent to learn a good policy for a difficult problem, with many suboptimal actions taken during the learning process.
When optimizing complex quantum systems, such as multi-level quantum systems, the complexity of the system dynamics increases sharply with the size of quantum systems \cite{niu2019universal}. Also, the dissipation part in a quantum system may irreversibly lead it away from an equilibrium state \cite{altafini2004coherent}, greatly increasing the difficulty of manipulating its dynamics. It is highly desirable to design an efficient DRL approach to achieve a fast and reliable control of complex quantum systems.

 Owing to the observation that students usually learn easy courses before they start to learn complex courses, curriculum learning \cite{bengio2009curriculum,narvekar2020curriculum} has emerged as a general and powerful tool for solving difficult problems and has also been applied in optimizing RL agents \cite{ren2018self,shao2018starcraft}. For example, the introduction of curriculum learning allows the RL agents to make the best use of transitions and finally achieves higher scores when playing with complex games \cite{ren2018self}.
Actually, curriculums can be defined in different levels, including ordering of tasks or ordering of individual samples \cite{ren2018self}. However,
creating a curriculum at the sample level can be computationally difficult for
large sets of tasks since the entire set of samples from a task (or multiple tasks) is typically not available ahead of time. In addition, the samples experienced in a task depend on the agent's behavior policy, which can be influenced by previous tasks. Therefore, a simplified representation of a curriculum is often adopted in practical applications to eliminate the need for the knowledge of all samples. 

In this paper, we introduce a task-level curricluum to the deep reinforcement learning control for quantum systems and propose a novel  curriculum-based deep reinforcement learning (CDRL) with the purpose of achieving reliable and fast manipulation of quantum systems.
In particular, a task is defined by a threshold of fidelity (defined as a target fidelity), which also represents the difficulty of one task. In particular, two methods of task generation for a curriculum are introduced, including presetting static fixed tasks using empirical knowledge or dynamically generating tasks based on the performance of the agent. 
By sequencing a set of tasks with increasing difficulties and reusing knowledge between different tasks, the RL agent is able to focus on easy tasks at the early stage, gradually improve its goal after grasping basic skills and therefore achieve the final task. 


In CDRL, a target fidelity is closely related to each task, which allows for an early termination of each episode once a satisfactory effect is achieved. Such a mechanism  makes it possible to manipulate quantum systems within shorter control pulses thus protecting systems against decoherence. This is particularly significant for open quantum systems, where dissipation typically drives the system to a mixed state and an optimal control might keep the system from the maximum entropy state for some specific dissipation channels \cite{lin2020time}.  From this respective, the proposed method is an attempt to utilize machine learning techniques to realize efficient manipulation of quantum systems, which not only saves operation time but also keeps systems away from unwanted decoherence \cite{berry2009transitionless}.



 In this paper, we focus on a basic and crucial issue of quantum state preparation, which aims at steering a quantum system from an initial state towards a target state. To test the effectiveness of the proposed method, numerical simulations on closed quantum systems and open quantum systems are implemented. The main contributions of this paper are summarized as follows.
\begin{itemize}
\item  A curriculum-based deep reinforcement learning method (CDRL) is proposed  for quantum systems where tasks among a curriculum are defined by fidelities and are linearly sequenced with increasing fidelities.
\item The CDRL method is applied to closed quantum systems and open quantum systems to achieve enhanced performance where open quantum systems suffer from undesirable dissipation effects.
\item The CDRL method has the advantage of searching for shorter control pulses, thus providing insights for utilizing machine learning techniques to achieve fast control of complex quantum systems. 
\end{itemize}


The rest of this paper is organized as follows. Section \ref{Sec:problem} introduces several basic concepts about RL, curriculum learning and quantum systems. In Section \ref{Sec:method}, the CDRL method is presented in detail  with curriculum design and algorithm implementation for quantum systems. Numerical results for both closed and open quantum systems are presented in Section \ref{Sec:simulation}. Concluding remarks are drawn in Section \ref{Sec:conclusion}.


\section{Preliminaries}\label{Sec:problem}

\subsection{Reinforcement Learning}\label{Subsec:rl}
 \emph{1) Markov Decision Process:} RL is commonly studied
based on the framework of \emph{Markov Decision Process} (MDP). An MDP can be described by a tuple
of $\langle S,A,P,R,\gamma \rangle$ \cite{sutton2018reinforcement}, where $S$ is the
state space, $A$ is the action space, $P$: $S\times A\times S \to
[0,1)$ is the state transition probability, $R$: $S\times A\to
\mathbb{R}$ is the reward function, and $\gamma\in[0,1]$
is the discount factor. A policy is a mapping
from the state space $S$ to the action space $A$. At each time step $t\in[0,T]$, where $T$ is the terminal time, the agent forms the state $s_{t}\in S$, takes an action $a_{t}\in A$ according to a certain policy $\pi$: $a_{t} = \pi(s_{t})$, transits to the next state $s_{t+1}$ and gets a scalar reward signal $r_{t}$ from the environment. RL aims at determining an
optimal action $a_t^{*}$ at each state $s_{t}$ so as to maximize the
cumulative discounted future rewards of return $R_t =
\sum_{k=0}^{T-t}\gamma^{k}r_{t+k}$.


\emph{2) Deep Q Network:}
Similar to Q-learning \cite{watkins1992q}, deep Q-network (DQN) is a value-based DRL method, and it employs a function with parameters $\xi$ to approximate Q-value for each state-action pair $Q(s,a)$, i.e., $Q(s,a;\xi) \approx Q(s,a)$. Such a network can be trained by minimising the loss function as:  $Loss(\theta) = [(y - Q(s,a;\xi^{-}))^{2}]$ with $y = r + \gamma\max\limits_{a^{\prime}}Q(s^{\prime},a^{\prime};\xi^{-})$ \cite{baird1995residual},
where $s^{\prime}$ is the next state after taking action $a$ at state $s$. $\xi^{-}$ denotes the parameters of the target network which is fixed during the computation of $y$ and is usually updated after
some training iterations. Differentiate the loss function with respect to $\xi$, the gradient is formulated as
\begin{equation}\label{eq:gradient}
\resizebox{.9\hsize}{!}{$\nabla_{\xi}Loss =[r + \gamma\max\limits_{a'}Q(s',a';\xi^{-})-Q(s,a;\xi)]\nabla_{\xi}Q(s,a;\xi)$}.
\end{equation}
After training the network, an optimal control strategy is generated by selecting the action with the largest Q-function, i.e., $a^{*}=Q(s,a)$.

\subsection{Curriculum learning}
Curriculum learning is a methodology to optimize the order in which experiences are accumulated by the agent \cite{narvekar2020curriculum}. During this process, knowledge is transferred from easy tasks to difficult ones, which helps achieve  enhanced performance on a hard problem or reduce the time it takes to converge to an optimal policy. A curriculum is commonly defined as an ordering of tasks; while an ordering of individual experience samples can also be regarded as a curriculum at a more fundamental level. In addition, a curriculum is not always a simple linear sequence. In fact, one task building upon knowledge gained from multiple source tasks is also acceptable and useful, which is similar to the case in human education, where courses can build off multiple prerequisites.

There have been growing interests in exploring how to leverage curriculum learning to speed up the RL agents. To achieve this, there are three key elements to be considered:
(i) Task generation: a set of intermediate tasks
can be pre-specified ahead of time or dynamically generated
based on the previous learning performance during the curriculum
construction. (ii) Sequencing: similar to task generation, it
can be predefined or automatically realized. (iii) Knowledge transfer: before moving to the next task, reusable
knowledge acquired from one task is required to extract and pass to the next one \cite{lazaric2008transfer,taylor2009transfer}.
From this respective, designing a good curriculum depends on generating appropriate and useful tasks to transfer or reuse knowledge between different tasks.



\subsection{Quantum dynamics}
The state of a finite-dimensional closed quantum system can be represented by a unit complex vector $|{\psi}\rangle$  and its dynamics can be described by the Schr\"{o}dinger equation:
\begin{equation}
\frac{d}{dt}|{\psi}(t)\rangle=-\frac{\rm{i}}{\hbar}(H_0+\sum_{m=1}^{M}u_m(t)H_m)|{\psi(t)}\rangle,
\label{eq:schron}
\end{equation}
where $\hbar$ is the reduced Planck constant (hereafter, we set $\hbar=1$), $H_0$ denotes the time-independent free Hamiltonian of the system, and the control Hamiltonian operators $H_m$ represent the interaction of the system with the control fields. To drive the quantum system from an initial state $|\psi(0)\rangle=|\psi_0\rangle$ to a target state $|\psi_f\rangle$ within a given time period $T$, we adopt the fidelity between the actual state $|\psi(T)\rangle$ and the target state $|\psi_f\rangle$, i.e., $J(u)= |\langle \psi(T)\rangle | \psi_f\rangle |^2$, to evaluate the control performance \cite{wiseman2009quantum}.

In practical applications, a quantum system usually suffers from the interaction with its environments and is then regarded as an open quantum one \cite{altafini2004coherent}. In such a case, the state of the quantum system is described by a Hermitian, positive semidefinite matrix $\rho$ satisfying $\textup{Tr}(\rho)=1$ and $\textup{Tr}(\rho^2)\leq 1$. Its dynamics under Markovian approximation can be described by the Lindblad master equation \cite{dong2010quantum}:
\begin{equation}
{\rm{i}}\dot{\rho}(t)=[H_0+\sum_{m=1}^{M}u_m(t)H_m,\rho(t)]+\sum_k \gamma_k \mathcal{D}[L_k](\rho(t)),
\label{eq:lindblad}
\end{equation}
with $\mathcal{D}[L_k](\rho)=L_k \rho L_k^{\dagger}-\frac{1}{2} L_k^{\dagger} L_k \rho-\frac{1}{2}\rho L_k^{\dagger} L_k$, where $L_k$ represents the Lindblad operators and the coefficients $\gamma_k\geq 0$ characterize the relaxation rates. For an  $n$-level quantum system, denote a set of orthogonal generators as $\{{\rm{i}}Y_l\}_{l=1}^{N}$ ($N=n^2$) of $\mathbf{SU}(\textit{n})$ with each generator satisfies (i) $\textup{tr}(Y_l)=0$, (ii) $\ Y_l^\dagger=Y_l$, (iii) $\textup{tr}(Y_l Y_m)=2\delta_{lm}$. Then, a coherent Bloch vector can be defined as $\textbf{y}(t):=(y_1(t),...,y_N(t))^\top$, $y_l(t)=\textup{tr}(Y_l \rho(t))$, which also means that a density operator can be rewritten as $\rho(t)=I/n+\frac{1}{2}\sum_{l=1}^{N}y_l(t) Y_l$. In that case, the evolution of $\textbf{y}(t)$ is deduced as
\begin{equation}
\dot{\textbf{y}}(t)=(\mathcal{L}_{H_0}+\mathcal{L}_D)\textbf{y}(t)+\sum_{m=1}^{M}u_m(t)\mathcal{L}_{H_m}\textbf{y}(t)+s_0,\quad\textbf{y}(0)=\textbf{y}_0,
\end{equation}
where $\mathcal{L}_{H_0}$, $\mathcal{L}_D$ and $\mathcal{L}_{H_m}$ are $N \times N$ superoperators and the  the inhomogeous cource term $s_0$ is a $N\times 1$ column vector \cite{yang2013exploring}. The goal is to steer the system from an initial state $ \textbf{y}_0 $ to a final state $ \textbf{y}(T)$ as close to a target state $ \textbf{y}_f$ as possible, and we can take the following cost function \cite{yang2013exploring}:
\begin{equation}
J(u)=1-\frac{n}{8(n-1)}\parallel\textbf{y}_f -\textbf{y}(T)\parallel^2.
\label{eq:open loss}
\end{equation}

\section{Curriculum-based deep reinforcement learning for quantum control}\label{Sec:method}
In this section, the framework of CDRL for quantum control is presented, and the key elements for designing a curriculum are elaborated in detail, and then the ingredients of applying DRL to quantum systems are provided. Finally, the implementation of CDRL for quantum control is summarized.

\subsection{Framework of CDRL}

\begin{figure*}
\centering
\includegraphics[width=0.9\textwidth]{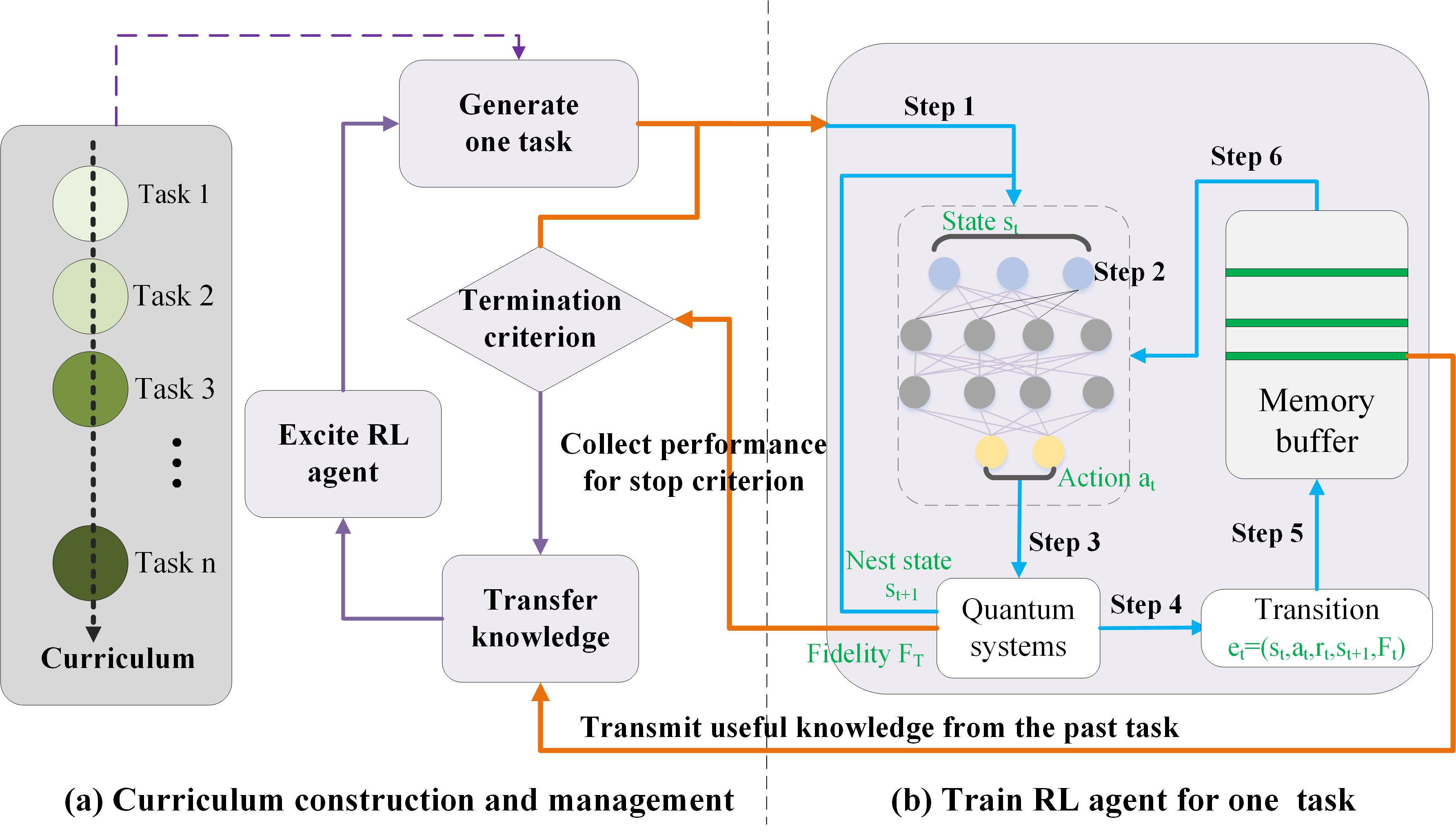}
\caption{Framework of CDRL approach for quantum control.
(a) Curriculum construction and management. The curriculum agent generates a task for training the RL agent and utilizes the performance of the RL agent to determine whether a stop criterion is satisfied. Once the previous task has been achieved, knowledge is transferred between different tasks by reusing past samples. Meanwhile, measures are taken to excite the RL agent before the subsequent task is scheduled.
(b) Train the RL agent for one task. At each time step $t$, the RL agent observes its current state $s_t$ (step 1) and
suggests a control action $a_t$ (step 2), which is mapped to the control
fields $\{u_m(t)\}$ (step 3). Then, the quantum system
takes the proposed control strategy and obtains the next state $s_t$, with fidelity $F_t$ and reward signal $r_t$ (step 4).
The transition $e_t=(s_t,a_t,r_t,s_{t+1},F_t)$ is put into a large memory
buffer (step 5). Finally, a batch of transitions is selected from the buffer and then fed into the networks to update its weights or parameters (step 6).}
\label{fig:framework}
\end{figure*}

Curriculum learning aims at generating appropriate and useful tasks and reusing knowledge between different tasks.  This allows the agent to focus on easy tasks at the early stage and gradually move onto difficult tasks. For each task, it is essential to train a DRL agent to achieve the task. The framework of  CDRL is presented in Fig. \ref{fig:framework}, which can be summarized as two aspects: (i) Curriculum construction and management; (ii) Train the DRL agent for one task.

In Fig. \ref{fig:framework}(a), a set of tasks is constructed and then compose a curriculum. During the learning process, once a task is generated, it is sent to the RL agent for learning. Meanwhile, the training performance collected from the RL agent is sent back to the curriculum agent for determining whether a stop criterion is met. Before moving to the next task, knowledge acquired from the previous task is collected and passed to the next one. In addition, some measures are taken to excite the RL agent to maintain a certain exploration for a new task.

In Fig. \ref{fig:framework}(b), the RL agent trains its network by trial-and-error for each task. After receiving a task, the RL agent observes its current state $s_t$ and determines an action $a_t$ by inferring from the deep neural networks. The control fields $u(t)=\{u_m(t), m=1,2,...,M\}$ based on $a_t$ are performed on the quantum system, resulting in a next state $s_{t+1}$ and fidelity $F_t$ and reward signal $r_t$. The transition $(s_t,a_t,r_t,s_{t+1},F_t)$ is collected and stored into a memory pool. Meanwhile, a batch of samples is selected from the memory and fed into the neural network to update its parameters. It is worthy to note that the experiences from the past task are collected as reusable knowledge for the next task.


\subsection{Curriculum design for quantum systems}

\subsubsection{Task and curriculum}

Most quantum control problems can be generalized as a goal task, which means that the agent starts from an initial state and attempts to approach a target state as close as possible with a given time period. In quantum information theory, the ``closeness" between two
quantum states is usually measured by fidelity. However, the ideal value of $F(|\psi(T)\rangle,|\psi_f\rangle)=1$ does not necessarily mean $|\psi(T)\rangle=|\psi_f\rangle$. In fact, there are multiple states $|\psi^{\prime}\rangle=\exp{(\rm{i}\theta)}|\psi(T)\rangle, \theta \in R$ that can achieve the same fidelity, i.e., $F(|\psi(T)\rangle,|\psi_f\rangle)=F(|\psi^{\prime}\rangle,|\psi_f\rangle)$. In that case, $|\psi^{\prime}\rangle$ is equivalent to $|\psi(T)\rangle$ neglecting the global phase. For ideal fidelity $1$, the final states are not unique, and the states in the last but one step can be multiple when tracing one step backward. From this respective, there exist multiple branches or trajectories that result in final states with the same fidelity. Hence, when designing a curriculum for quantum systems, we do not
take the actual physical states as the intermediate tasks. Instead, a target fidelity is utilized to distinguish between different tasks. Based on that, we have the following definition.

\begin{definition}[Task]
For a quantum system with a target state $|\psi_f\rangle $, one task $v$ is defined as the process of driving the system from an initial state $|\psi_0\rangle$ to an actual state $|\psi(T)\rangle$ with $F(|\psi(T)\rangle,|\psi_f\rangle) \geq D(v)$. $D(v)$ represents the difficulty of the task $v$.
\end{definition}

Recall that a curriculum is composed of a set of tasks. The ordering of tasks among a curriculum is similar to the way that vertices and edges form a graph. Hence, we adopt the concept of graph to define a curriculum.  In particular, a task is defined as a vertice of a graph, and the relationship between two successive tasks are used to define an edge of a graph. 

To achieve a final goal, it is natural to sequence the tasks according to their difficulty. Denote the $i$-th task as a vertice $v_i$. Then a directed edge $<v_i,v_j>$ can be utilized to denote the relationship between two tasks, which means that samples associated with $v_i$ should be trained before samples associated with $v_j$.
By sequencing a set of different tasks in a similar fashion, a task-level curriculum can be in the following way:
\begin{definition}[Curriculum]
Let the task set be $V$ and samples relating to $V$ be $\mathbb{M}^{V}$.
A curriculum can be defined as a directed acyclic graph $C = (V,\varepsilon,g)$, where $V$ is the set of vertices, $\varepsilon\subseteq \{(v_i,v_j) | (v_i,v_j)\in V \times V \bigwedge D(v_i) < D(v_j)\}$ is the set of directed edges, and $g: V \rightarrow \mathbb{P}(\mathbb{M}^{V})$ is a function that associates vertices to subsets of samples in $\mathbb{M}^{V}$, where $\mathbb{P}(\mathbb{M}^{V}) $ is the power set
of $\mathbb{M}^{V}$.
\end{definition}

\begin{figure}
	\centering
	\includegraphics[width=0.45\textwidth]{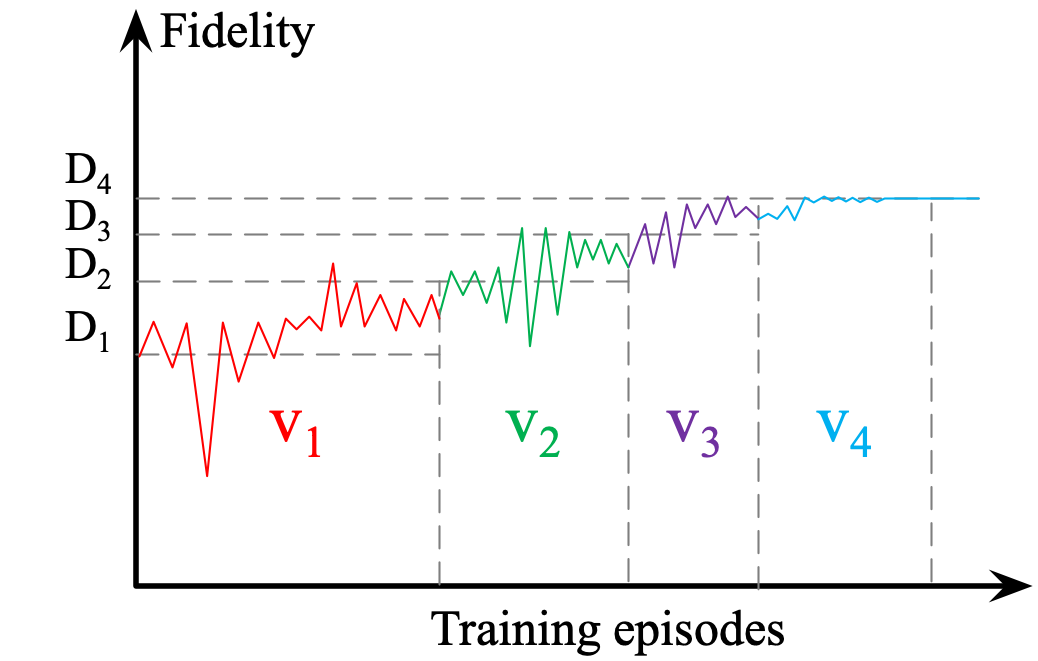}
	\caption{Illustration of the curriculum learning process with a flow of tasks. }
	\label{fig:multi-stage DRL}
\end{figure}

Based on the definition of tasks, a curriculum for quantum problems can be simplified by sequencing tasks in a linear way. In that case, the graph is reduced to a chain, where the indegree and outdegree in the graph $C$ are at most 1.  Following the linear chain of a curriculum,  the whole learning can be described as Fig.  \ref{fig:multi-stage DRL}, where the learning process is in line with a flow of tasks with increasing difficulties. For each task $v_i$, the agent tries to achieve a better performance than the target fidelity $D(v_i)$. After learning for some episodes, the agent moves on to the next task $v_{i+1}$. Although there are oscillations during each task, the average performance reveals that the agent has climbed to a higher point.


\subsubsection{Task generation for a curriculum}

During the construction of a curriculum, generating useful and appropriate tasks is a crucial procedure \cite{narvekar2016source,narvekar2017autonomous} since the quantity of the created tasks has a strong impact on the search space and efficiency of curriculum sequencing algorithms. Therefore, all the
generated tasks should be relevant to the final task(s). A common choice is to manually design a set of intermediate tasks such that  transferring knowledge through them is beneficial. This is usually realized based on the empirical knowledge about the problem. Recently, there are efforts to automatically create tasks online using a set of rules or a generative process \cite{narvekar2017autonomous}, even though these approaches may still rely on tuning hyper-parameters, such as the number of tasks generated.


Recalling that a task for quantum systems is determined by a fidelity.
In principle, it can achieve any value that lies between 0 and 1 when arbitrary control laws are available. Hence, any value lies in $[0,1]$ can be utilized as an indicator for one task. Moreover, tasks are usually ordered with increasing difficulty. Based on this knowledge, the preset tasks should be arranged in such a way that the front vertices are assigned with low target fidelities; while those vertices at the rear are attached with difficulties approaching $1$. Moreover, the difficulty of controlling quantum systems increases tremendously. When determining a set of tasks with increasing difficulties, the gap between two successive tasks $\delta_i=D(v_{i+1})-D(v_{i})$ should be lowered down with increasing $i$.

Another way is to dynamically generate tasks during the construction of a curriculum. To start a curriculum, an initial task should be set first. A practical way is to preset an appropriate initial value which is easy to achieve without learning. Such a value can be given based on prior knowledge, or by observing some random trials of the RL agents. For the left tasks, it is important to perform data statistics on the performance for the previous task to suggest a candidate fidelity for the next task. For example, it is useful to measure the ``mean/median/max" of the fidelities during the past episodes. Likewise, the determination of the next task should follow the principle that the new task is assigned with higher difficulty compared with the previous task.



\subsection{CDRL for quantum control}

\subsubsection{DRL for quantum control}


When applying RL to quantum systems, a maximum number of control pieces is usually defined as $N_{\max}$, which limits the maximum steps in one episode. This is consistent with traditional learning methods such as gradient method (GD) and genetic algorithm (GA), where the performance of one trial is evaluated after all pieces of control fields have been performed. In RL, the control performance is evaluated step by step, and the reward signal at the current step is sent to the agent to suggest the control field for the next step. It is possible that in one episode, the performance does not always increase with steps and may decrease to a lower value after performing additional control pulses, especially when the previous step has achieved excellent performance. Hence, we take a smart solution in this approach, which is to terminate the current episode when certain requirements are satisfied \cite{mackeprang2019reinforcement,zhang2019does}. 
Recall that the RL agent usually takes some episodes to realize each subtask. We associate the two factors together and introduce a new definition for one episode.


\begin{definition}[Smart Episode]
For an RL agent aiming at achieving task $v$, a smart episode is defined as a state-action chain $s_0 \xrightarrow {a_0} s_1 \xrightarrow { a_1} ... s_j \xrightarrow {a_j} s_{j+1}... $ with termination conditions as (i) the number of steps taken reaches the maximum steps, i.e., $j=N_{\max}$; or (ii) the current performance surpasses the target fidelity, i.e., $F(|\psi_{j+1}\rangle,|\psi_f\rangle)\geq D(v)$.
\end{definition}
From this respective, the number of control pieces in the DRL approach can be less than $N_{\max}$. 
This measure enables  the RL agent to search for shorter control pulses.  For convenience, we assume an equal time duration for each piece, denoted as $dt$ and a non-negative integer $j$ is used to indicate the time of $jdt$. For example, we have $|\psi_j\rangle = |\psi(jdt)\rangle$ and $F_j=F(|\psi_{j+1}\rangle,|\psi_f\rangle)$.

Another significant procedure for RL methods is to determine an appropriate reward signal \cite{sutton2018reinforcement}.
Recalling that, the performance in one episode is not necessarily increasing, the reward signal is only given at the end of the episode. In this work, the reward is calculated based on the infidelity and the function $\log_{10}(\cdot)$ to attach higher reward signal to higher fidelity. Finally, the reward signal can be formulated as
\begin{equation}
\resizebox{.95\hsize}{!}{$
r_j = \left\{\begin{array}{ll} k_1+k_2\log_{10}(1-F_j) & \textup{if} \quad F_j>D(v) \quad \textup{or} \quad j=N_{max} \\
0  & \textup{else}
\end{array}\right.
$}
\label{eq:rewardpolicy}
\end{equation}
where $k_2$ represents the slope of reward function, and $k_1$ acts as a bias. Actually, $k_1$ and $k_2$ can be adjusted according to the different fidelity intervals \cite{zhang2019does}.
In principle, a large value of $k_2$ should be set for a large fidelity such as $[0.99,0.999]$, while the value of $k_1$ guarantees that reward increases with fidelity between different intervals.


When training DRL agents, experience replay \cite{lin1993reinforcement} is utilized to store the agents' past experiences into a big memory pool for replaying \cite{an2019deep}. Considering that transitions may be more or less surprising, redundant, or task-relevant. When attempting to achieve a certain task, we employ a smart store mechanism to make better use of significant samples. For each step,  the fidelity between $s_{j+1}$ and the target state is informative, and we define a transition as $e_j=(s_j,a_j,r_j,s_{j+1},F_j)$.  If the older one has higher fidelity than the new one, the new sample should be moved onto the next pointer for replacement and storing. This helps retain some significant samples while storing the latest experiences into the memory. It is worthy to note, such a practice is based on the hypothesis that good transitions are usually not ordered nearby and the next sample after a good sample is rarely a good one owing to the smart episode. Besides, this measure only takes effect with a certain probability.

\subsubsection{Knowledge transfer between different tasks}
\begin{figure}
	\centering
	\includegraphics[width=0.45\textwidth]{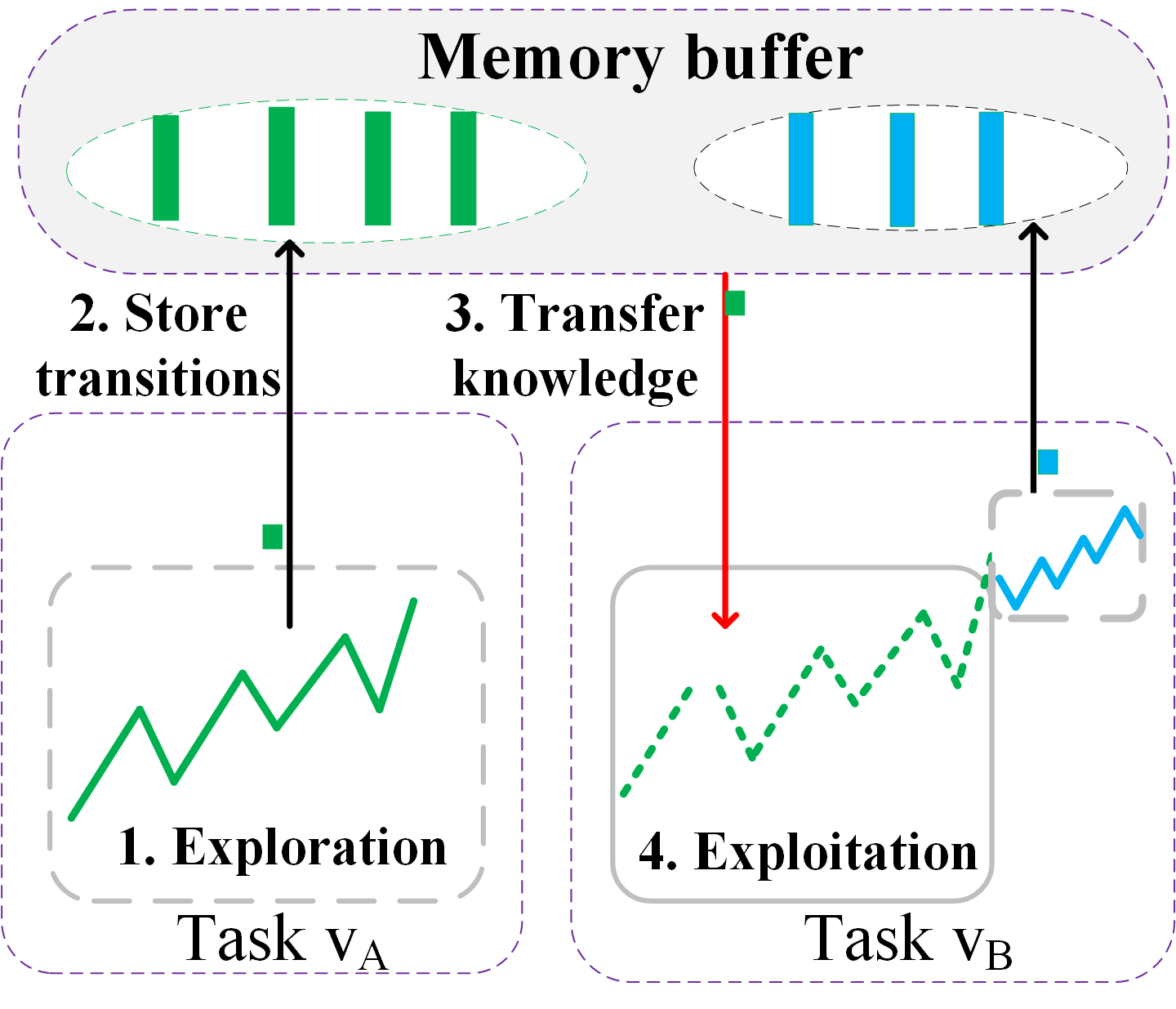}
	\caption{Illustration of transferring knowledge between two tasks.}
	\label{fig:knowledgetransfer}
\end{figure}

After training the DRL agent for each task, it is important to transfer and reuse knowledge among different tasks. Given two tasks, the process of transferring knowledge can be summarized as four procedures in Fig. \ref{fig:knowledgetransfer}.

(i) Exploration: On the previous task $v_A$, the agent attempts to achieve a higher fidelity than $D(v_A)$. After many trials, the agent acquires the knowledge by updating its weights using the useful transitions during task $v_A$. (ii) Storing transitions: Meanwhile, those useful transitions are collected and stored in the memory pool. (iii) Transferring knowledge: Those transitions from task $v_A$ provide useful resources and can be replayed to train the agent when striving for the subsequent tasks. (iv) Exploitation: When attempting to achieve task $B$, the agent already has the capacity to achieve a fidelity $F(|\psi\rangle,|\psi_f\rangle) >D(v_A)$ within $n_a$ steps by repeating the past transitions. Continuing that trajectory, the agent attempts to search another $\delta_n$ steps with final state $|\psi^{\prime}\rangle$, hopefully to achieve a better fidelity $F(|\psi^{\prime}\rangle,|\psi_f\rangle) > D(v_B)$. This guarantees that the agent reviews the knowledge accumulated from task $v_A$ and explores new transitions under task $v_B$. From this respective, the process for transferring knowledge between different tasks is a combination process of exploitation and exploration.


\subsubsection{Stop criterion and excitation}
To achieve a better balance between exploration and exploitation for curriculum learning, it is important to set a suitable stopping criterion for each task. Typically, training is terminated when performance on a task or a set of samples has converged. However, training to convergence on an intermediate task is not always necessary since the convergence of DRL is not available and is difficult to evaluate \cite{narvekar2020curriculum}. Another option is to train on each task for a fixed number of episodes. 

In practical applications, the performance of different episodes tends to vary frequently since the randomness or uncertainties in actions and the stochastic nature of batch sampling for updating parameters might result in fluctuation of the DRL agent. When dealing with quantum systems, the subsequent state varies greatly when taking one step following different actions, thus leading to great difference in the final performance. In that case, a fixed number of episodes may not be enough to guarantee good performance on one task. In addition, the process of winning a task may take more episodes for a harder task. Based on the above observations, we propose a novel stop criterion. For task $v_i$, a non-negative integer $w$ is introduced to measure the time of hitting the task, that is to update $w \leftarrow w+1$ once the current episode achieves $F_t \geq D(v_i)$ with $t\leq N_{\max}$. Besides, an integer $SC$ is defined to terminate the current task when $w \geq SC$. 

During each task, the randomness factor for the DRL agent is usually decreased using a decay factor $\lambda \in(0,1)$ to guarantee a certain convergence. For example, the greedy factor $\epsilon$ for discrete control cases tends to reach 0, or the disturbance amplitude for continuous control cases approches nearly 0. To achieve a smooth transition between two tasks, an excitation operation is required to reset the randomness degree of the DRL agent to guarantee a certain exploration for the new task.


\subsection{Integrated algorithm}

\begin{algorithm*}[ht]
\caption{Algorithm description for CDRL}\label{al:CDRL}

\KwIn{Initial state $|\psi_0\rangle$, target state $|\psi_f\rangle$, maximum steps $N_{\max}$, decay period $E$, $\epsilon$-decay factor $\lambda$,
 \newline batch size $B$, memory size $K$, count threshold $SC$, probability threshold $\mu_1$}
\KwOut{Optimal set of control pulses $a_j=\max_{a} Q(s_j,a)$}

Build a DQN agent $Q(s,a|\xi)$ and initialize $i=0$, $k=0$ and a memory pool $\mathcal{M}$\;
Initialize \emph{task} $v_0$ with an initial target fidelity $D(v_0)$\;

  \While{ curriculum not finished}
  {
   Initialize the number of hitting a task as $w=0$ and a null list $L$ to store the past performance\;

   \While{ stop criterion for \emph{task} $v_i$ not met}
  {
    Initialize step $j=0$ and transform initial quantum state $|\psi_0\rangle$ into $s_0$\;
   \While{$s_j$ is not termination state}
  {
    Generate actions $a_j$ based on $Q(s_j,a_j|\xi)$ following the $\epsilon$-greedy action selection policy\;
    Perform the control pulses obtained from $a_j$, obtain next state $s_{j+1}$ and fidelity $F_j=F(|\psi_{j+1}\rangle,|\psi_f\rangle)$\;
    Determine $r_{j}$ based on $F_j$ and $D(v_i)$\;

      \If{$F(|\psi_j\rangle,|\psi_f\rangle) \geq D({v_i})$ or $j\geq N_{\max}$ }
   { $s_j$ is termination state;}
   { $j \leftarrow j+1$ }

   \If{Memory pool is full}
   {
    Sample a batch of samples and update parameters $\xi$ according to Eq. (\ref{eq:gradient})\;
    Observe the fidelity of $k$-th sample in $\mathcal{M}$, denoted as $h(k)$\;
    \If{ $ h(k)> F_j$ \textup{and} $\textup{rand}(1) < \mu_1$ }
    {
    $k \leftarrow k+1 $\;  
    }
   }
   Store transition $e_j=(s_j,a_j,r_j,s_{j+1},F_j)$ as the $k$-th sample into $\mathcal{M}$ and set $k \leftarrow k+1 $\;
   }
   Decrease $\epsilon$ every $E$ episodes as $(1-\epsilon) \leftarrow \frac{(1-\epsilon)} {\lambda}$\;
   Label the performance of current episode as $F^*=F_j$ and append $F^*$ into $L$\;

  \If {$F^* \geq D(v_i)$}
      {Update the number of hitting current task $w \leftarrow w+1$\;}
  \If{$w > SC $}
  {
   Stop criterion for the current task is met\;
  }

  }
   Determine the candidate difficulty based on $L$ and assign it to the next task \emph{task} $v_{i+1}$\;
   Reactivate DQN agent and reset $\epsilon$ to guarantee exploration\;
  $i \leftarrow i+1$

  }

\end{algorithm*}

For the implementation of CDRL for quantum control, recalling that task generation lies in the heart of curriculum construction, we take the approach of dynamically generating tasks as an example. Besides, an effective DRL method is required to generate an optimal policy for the agent.
Considering the wide application of DQN \cite{mnih2015human}, we adopt DQN as the baseline algorithm to train the agent towards a subtask. 

When searching for the optimal control fields, the set of $2^{M}$ possible choices for the action vector are formulated as \cite{an2019deep}
\begin{equation}
\mathcal{A}(u)=\{(u_1,...,u_M), u_m \in \{-G_m,G_m\},m=1,...,M\},
\label{eq: Action}
\end{equation}
where $\{G_m>0\}$ are preset bounds for each control. During the traning process, the Q-values for all possible actions, i.e., $Q(s,a_1), Q(s,a_2), ...,$ are predicted and the actual action is obtained in an $\epsilon$-greedy way. That is to select a random action $a_t$ with probability $\epsilon$ or to select $a_j = \max_a Q(s_j,a)$ with probability $(1-\epsilon)$. The randomness factor $\epsilon$ reflects the greedy degree, which means that a high value favors exploration while a low value encourages exploitation.

Finally, an integrated CDRL algorithm using dynamically generated tasks and DQN baseline is summarized as in Algorithm
\ref{al:CDRL}. The curriculum agent launches a task $v_i$, with an appropriate difficulty $D(v_i)$, to activate the DRL  agent to learn a good policy through trial-and-error. During each step, the agent
encounters one transition $(s_j,a_j,r_j,s_{j+1},F_j)$. By evaluating $F_j$, the transition $e_j$ is stored in the memory pool in the right order. Meanwhile, the DQN agent updates its weights by sampling a batch of those transitions. During this time, the performance of each episode is collected and transmitted to the agent for the determination of the stop criterion. Once it is met, the curriculum agent schedules a new task based on the past performance. This procedure is carried out iteratively until the final task is achieved.


\begin{remark}
In practical implementation, the gap of difficulty between successive tasks, i.e., $\delta_{i}=D(v_{i+1})-D(v_{i})$, should be evaluated to determine whether or not to terminate the learning process. For example,  if $\delta_{i}$ is below a small threshold (such as 0.00005), there is no need to introduce additional tasks, since the actual performance of one episode can be in principle larger than $D(v_i)$. As for the case of manually generated tasks, the agent may fail to achieve a better performance regarding a difficult task. To overcome this problem, a careful check is required to judge whether or not the current task is too hard for the agent. For example, the agent is thought to reach its final task if it fails to hit the task in the early episodes.
\end{remark}





\section{Numerical simulations}\label{Sec:simulation}

To test the performance of the proposed CDRL algorithm for manipulating quantum systems, several groups of numerical simulations are carried out for closed and open quantum systems, and the results are demonstrated and analyzed with detailed parameter settings. 

\subsection{Parameter settings }\label{Subsec:parameter}

Recall that there are two ways to generate tasks for a curriculum. 
For static one, several benchmark values are usually selected based on empirical knowledge. For example, it is appropriate to initialize the first task with difficulty $0.9$, while $0.99$ means a higher requirement and can be assigned to another task. Generally, the ability to achieve $0.999$ usually means an ideal case, and therefore a final task with difficulty $0.999$ can be obtained. Since a direct increase from $0.9$ to $0.99$ may not be achieved in the learning process, finely grained transitions between two tasks are required to generate more tasks. In the simulations, an incremental value between $0.9$ and $0.99$ is set as 0.02 and 0.01 for 2-qubit and 3-qubit systems, respectively; while an incremental value of 0.001 between 0.99 and 0.999 is adopted.

As for dynamic task generation, we initialize the first task with fixed target fidelity of $0.9$. For the left tasks, we take the median value of the fidelities collected during the previous task since a median value reflects the average performance and is robust to extreme values among the data.  


For reward singnal,  we design four piecewise functions to indicate different rewarding schemes. The parameters for $k_1$ and $k_2$ in (\ref{eq:rewardpolicy}) are given as follows:
\begin{equation}
(k_1,k_2)=\left\{\begin{array}{ll} (0,-10)& F_j \in (0,0.9)  \\
(60,-10)& F_j \in [0.9,0.99) \\
(-10,-100) & F_j \in [0.99,0.999) \\
(-800,-400) & F_j \in [0.999,1.000)
\end{array}\right.
\label{eq:reward}
\nonumber
\end{equation}
The discount factor is set as $\gamma=0.95$, and the $\epsilon$-greedy factor is initialized as $\epsilon=0.2$. The trained policy of DRL is approximated by a deep neural network with two 256-unit hidden layers. Parameters are optimized using adam with the learning rate of $\alpha=0.0001$. The decay period is set as $E=10$ and the batch size is set as $B=128$. The threshold for the stop criterion of each task is set as $SC=2000$. The above parameters are used in all simulations.  The other parameters in different simulations are summarized as in Table \ref{tab:paraDRL}. It is worthy to note that a prioritized experience replay method \cite{schaul2015prioritized} is applied and the probability threshold for the smart storing mechanism is set as $\mu_1=0.8$.

\begin{table}[!htbp]
	\renewcommand\arraystretch{2}
	\centering \caption{hiper parameters} \label{tab:paraDRL} \scalebox{0.95}{
		\begin{tabular}{c|c|c|c|c}
			\hline
			Parameters & 2-qubit & 3-qubit & 3-level open & 4-level open\\
			\hline
			Memory size & 20000  & 200000 & 20000 & 100000\\
			\hline
			Replace iteration  & 100 & 500 & 100 & 300\\
			\hline
			$\epsilon$-decay factor & 0.999 & 0.9995 & 0.999 & 0.999\\
			\hline
	\end{tabular}}
	\label{tab:parameters}
\end{table}

To verify the effectiveness of CDRL, we also compare CDRL with two traditional DRL versions. (i) DRL-1 searches the exact $N_{\max}$ pulses of control for each episode. (ii) DRL-2 takes an empirical target fidelity to terminate an episode. To achieve a valid comparison, DRL-1 and DRL-2 employ DQN with the same experience replay method as the baseline and adopt the same parameter settings as CDRL. For DRL-2, the target fidelity is set as 0.999 and 0.99 for 2-qubit and 3-qubit systems, respectively. For open quantum systems, it takes a value of 0.99 and 0.97 for three-level and four-level quantum systems, respectively. 

In addition, we also include the simulation of GD and GA for comparison. Their algorithm description and parameter settings are given in Appendix \ref{app:GDsetting}. 
In this paper, the simulations for each task are run multiple times and a seed is utilized to control the randomness for each running. The simulations can be  implemented in two scenarios according to the initial states: (i) a benchmark case means 10 runs with an identical initial state and different seeds, (ii) a random case means 10 runs with identical seed and different initial states. 


\subsection{Numercial results for closed quantum systems}\label{Sec:closed}

To apply DRL methods to quantum systems, we firstly build a map between their concepts. Considering that $|\psi_j\rangle$ is usually represented in a complex vector, a map is utilized to transform a quantum state to a real vector, which can be formulated as
\begin{equation}
s_j=[\textup{Re}(\langle 0 |\psi_j\rangle),\textup{Im}(\langle 0 |\psi_j\rangle),...,\langle n-1 |\psi_j\rangle),\textup{Im}(\langle n-1 |\psi_j\rangle)],
\end{equation}
where $\{|k\rangle\}_{k=0}^{n-1}$ is a set of basis states for an $n$-level quantum system. $\textup{Re}(·)$ and $\textup{Im}(·)$ are the functions of taking the real and imaginary parts of a complex number, respectively. 

\subsubsection{2-qubit quantum systems}\label{Sec:system}

\begin{figure*}[ht]
\centering
\begin{minipage}{0.25\linewidth}
  \centerline{\includegraphics[width=2in]{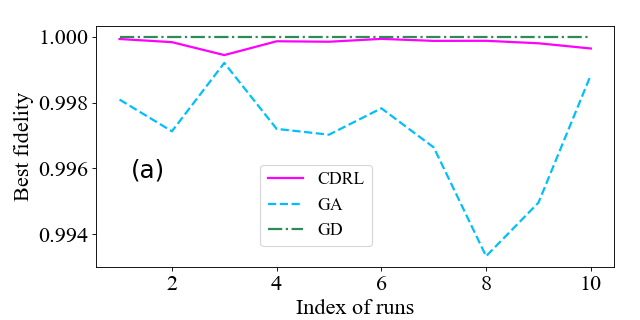}}
\end{minipage}
\hspace{1.2cm}
\begin{minipage}{0.25\linewidth}
  \centerline{\includegraphics[width=2in]{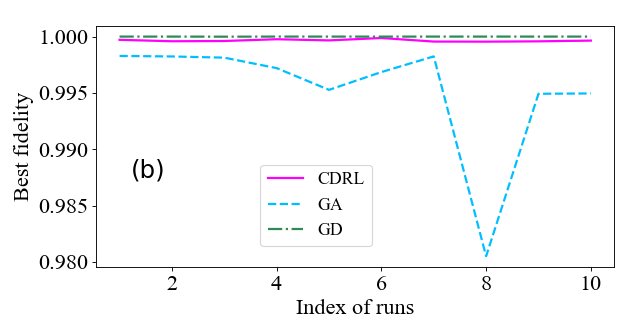}}
\end{minipage}
\hspace{1.2cm}
\begin{minipage}{0.25\linewidth}
  \centerline{\includegraphics[width=2in]{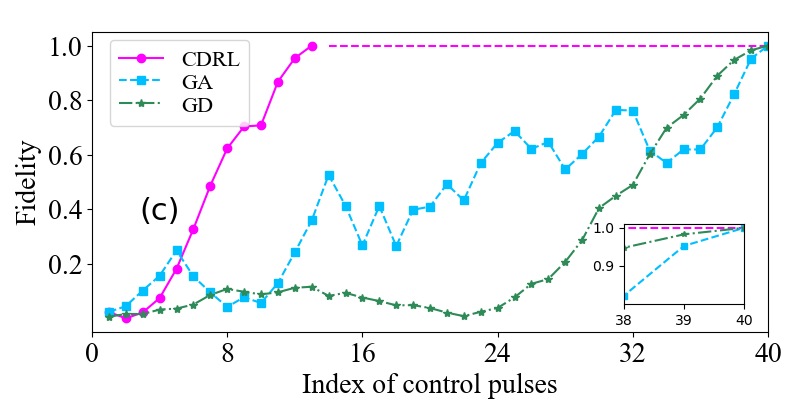}}
\end{minipage}
\caption{Comparison between GD, GA, CDRL on 2-qubit quantum systems. (a) Benchmark case, (b) Random case, (c) An example of trajectories.}
\label{fig:level4EAs}
\end{figure*}

\begin{figure*}[ht]
\centering
\begin{minipage}{0.25\linewidth}
  \centerline{\includegraphics[width=2in]{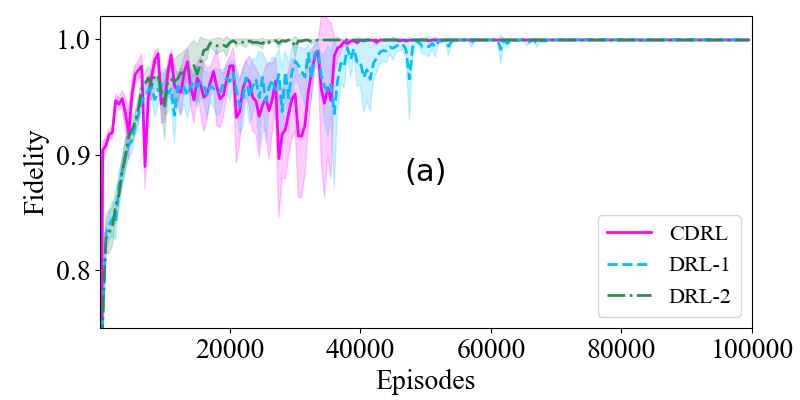}}
\end{minipage}
\hspace{1.2cm}
\begin{minipage}{0.25\linewidth}
  \centerline{\includegraphics[width=2in]{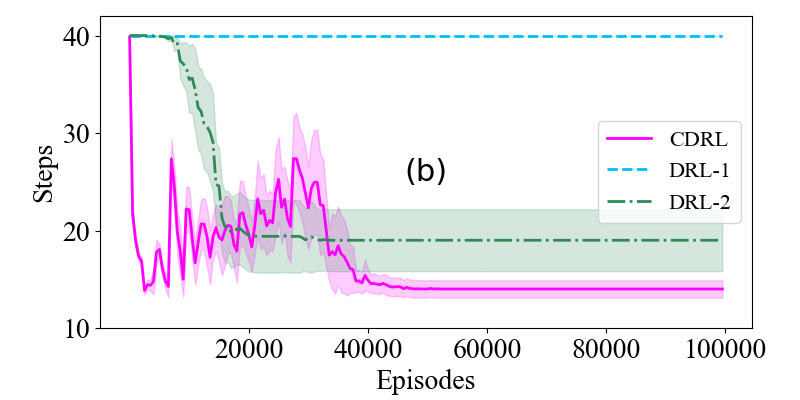}}
\end{minipage}
\hspace{1.2cm}
\begin{minipage}{0.25\linewidth}
  \centerline{\includegraphics[width=2in]{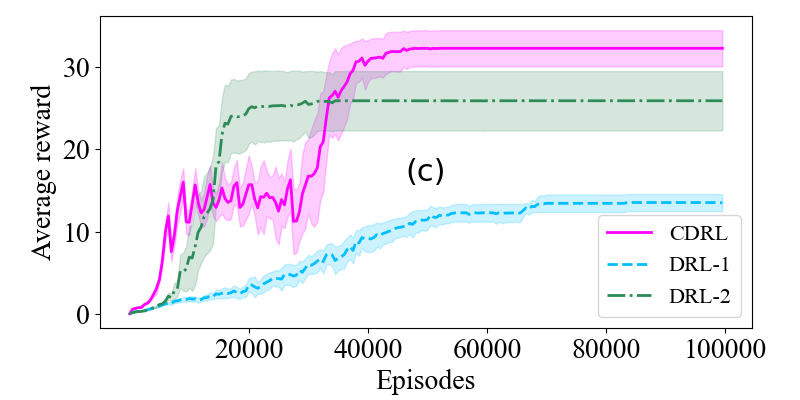}}
\end{minipage}
\caption{Comparison between different DRL approaches on 2-qubit systems for benchmark case. (a) Fidelity at the end of one episode, (b) Steps in one episode, (c) Average reward for one episode.}
\label{fig:close2qubit-bench-DRL}
\end{figure*}

\begin{figure*}[ht]
	\centering
	\begin{minipage}{0.25\linewidth}
		\centerline{\includegraphics[width=2in]{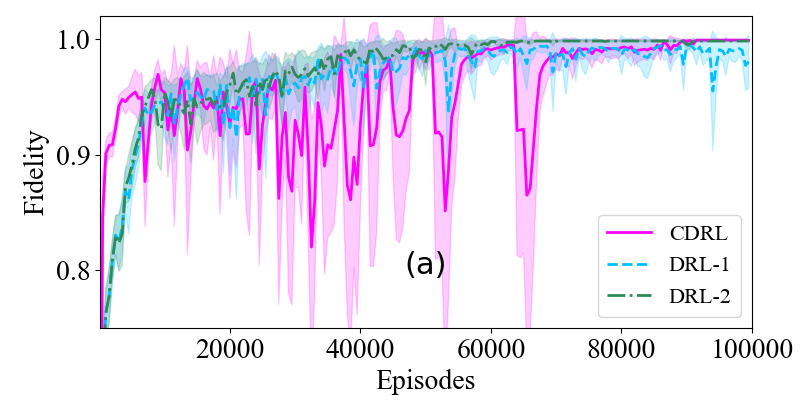}}
	\end{minipage}
	\hspace{1.2cm}
	\begin{minipage}{0.25\linewidth}
		\centerline{\includegraphics[width=2in]{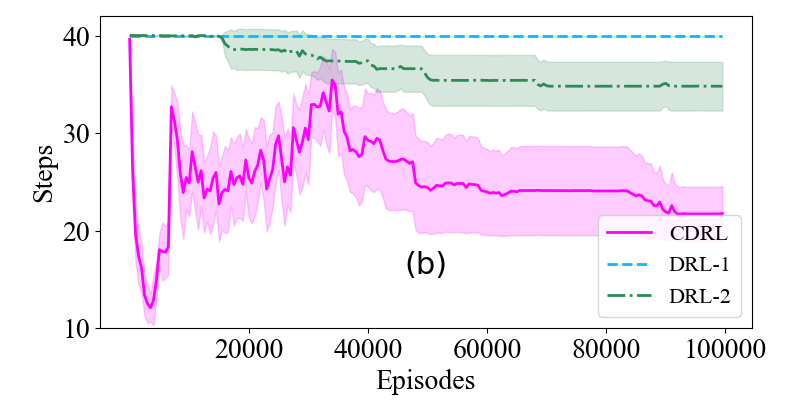}}
	\end{minipage}
	\hspace{1.2cm}
	\begin{minipage}{0.25\linewidth}
		\centerline{\includegraphics[width=2in]{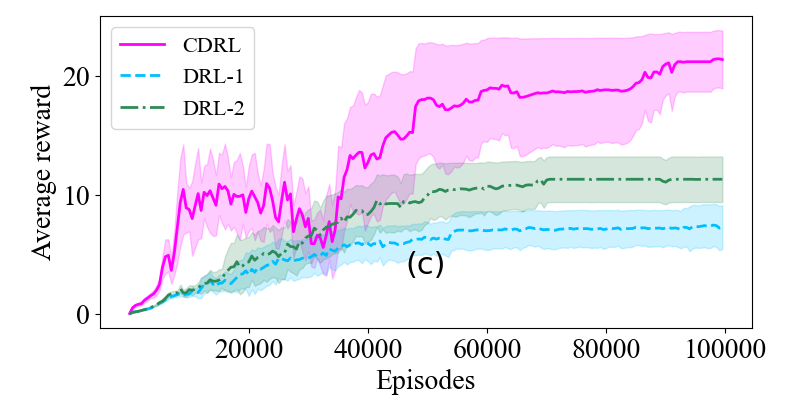}}
	\end{minipage}
	\caption{Comparison between different DRL approaches on 2-qubit systems for random case. (a) Fidelity at the end of one episode, (b) Steps in one episode, (c) Average reward for one episode.}
	\label{fig:close2-qubit-random-DRL}
\end{figure*}

\begin{figure*}[ht]
\centering
\begin{minipage}{0.25\linewidth}
  \centerline{\includegraphics[width=2in]{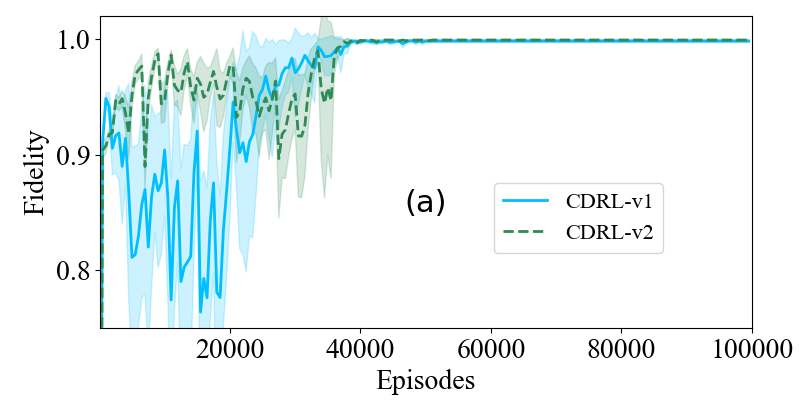}}
\end{minipage}
\hspace{1.2cm}
\begin{minipage}{0.25\linewidth}
  \centerline{\includegraphics[width=2in]{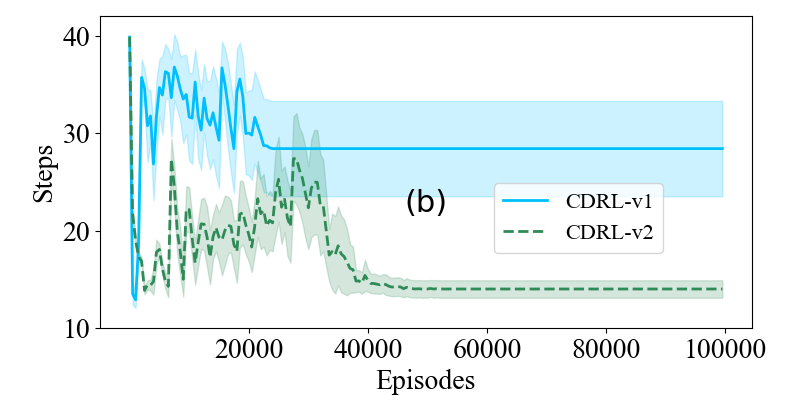}}
\end{minipage}
\hspace{1.2cm}
\begin{minipage}{0.25\linewidth}
  \centerline{\includegraphics[width=2in]{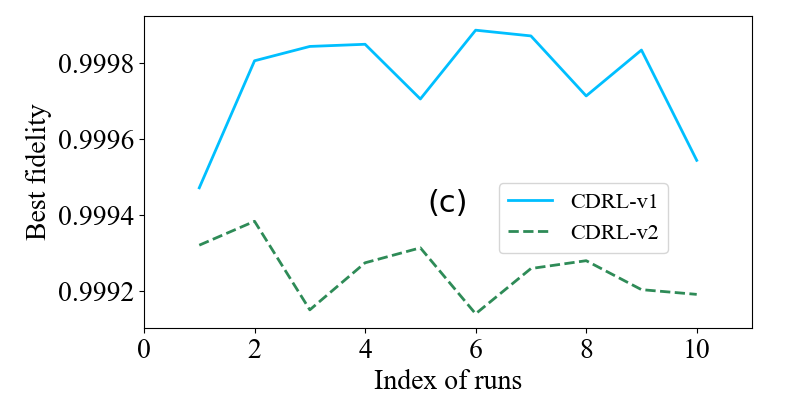}}
\end{minipage}
\caption{Comparison of two CDRL versions with different parameter settings. (a) Fidelity at the end of one episode, (b) Steps in one episode, (c) The best fidelities for 10 runnings.}
\label{fig:CDRLtwoversion}
\end{figure*}

Consider a 2-qubit system with Hamiltonian $
H(t)=H_0+\sum_{m=1}^{4}u_m(t)H_m$. Denote $I_n$ as the identity operator for $n$-level systems.  Denote the Pauli operators as
\begin{equation}\label{eq:Pauli operators}
\sigma_{x}=\begin{pmatrix}
0 & 1  \\
1 & 0  \\
\end{pmatrix} , \ \ \ \
\sigma_{y}=\begin{pmatrix}
0 & -i  \\
i & 0  \\
\end{pmatrix} , \ \ \ \
\sigma_{z}=\begin{pmatrix}
1 & 0  \\
0 & -1  \\
\end{pmatrix}. 
\end{equation}
We assume the free Hamiltonian is $H_0=\sigma_{z}\otimes \sigma_{z}$ and
the control Hamiltonian operators are $H_1=\sigma_{x}\otimes I_2$, $H_2= I_2 \otimes \sigma_{x}$, $H_3=\sigma_{y}\otimes
I_2$, $H_4= I_2 \otimes \sigma_{y}$, respectively. The maximum control step is defined as $N_{\max}=40$ and the control magnitudes are constrained as $G_1=G_2=G_3=G_4=4$.

Here, we take a static method to generate tasks for curriculum construction. The numerical comparison of CDRL, GD, and GA under $dt=0.0275$ is revealed in Fig. \ref{fig:level4EAs}. From Fig. \ref{fig:level4EAs}(a) and Fig. \ref{fig:level4EAs}(b), it is clear that GD achieves the best fidelities, closely followed by CDRL in both benchmark and random scenarios. In addition, CDRL achieves better fidelities than GA in searching for discrete control fields. The trajectories of the optimal control pulses learned by three methods are depicted in Fig. \ref{fig:level4EAs}(c), where CDRL finds an optimal control strategy with 13 control pulses to achieve a final fidelity of 0.9999. While the optimal control pulses searched by GA and GD take the maximum steps. In particular, the fidelity of CDRL increases with step, while the other two methods do not have this benefit.  In this sense, CDRL method has an advantage to search shorter control pulses without sacrificing its fidelity.  

The comparison between CDRL, DRL-1 and DRL-2 for the benchmark case is summarized as in Fig. \ref{fig:close2qubit-bench-DRL}. As we can see, CDRL converges to a similar fidelity with comparison to DRL-1 and DRL-2, although they display a little difference in the early stage in Fig. \ref{fig:close2qubit-bench-DRL}(a). From Fig. \ref{fig:close2qubit-bench-DRL}(b), the average steps learnt by CDRL are much lower than the other two methods. In Fig. \ref{fig:close2qubit-bench-DRL}(c), CDRL ranks first regarding the average reward in an episode (defined as $\frac{1}{N}\sum_{j=0}^{N} r_j$). For the random case in Fig. \ref{fig:close2-qubit-random-DRL}, CDRL tends to achieve a little better fidelity, as the violet line is higher than the other two lines. Similar to the benchmark case, CDRL also converges to a small value of steps compared with the other two DRL methods. It is worthy to note that the steps in an episode actually means the number of control pulses for a control strategy. Based on the above analysis, the convergence to shorter steps (in an episode) reveals that CDRL has the potential to search for shorter control pulses compared to DRL-1 and DRL-2.

Recall that success count $SC$ and decay factor $\lambda$ are two important factors for CDRL, we compare two versions of CDRL with different parameter settings: (i) CDRL-v1 with $SC=500$ and $\lambda=0.99$ and (ii) CRDL-v2 with $SC=2000$ and $\lambda=0.999$. Their comparison results are summarized as in Fig. \ref{fig:CDRLtwoversion}, where they converge to similar fidelities in Fig. \ref{fig:CDRLtwoversion}(a), but exhibit different performance in steps in Fig. \ref{fig:CDRLtwoversion}(b). Regarding the best fidelity, CDRL-v1 is shown to achieve superior performance over CDRL-v2 in Fig. \ref{fig:CDRLtwoversion}(c). The above results reveal that CDRL with a longer learning period for each task tends to converge to smaller steps, while a shorter learning period helps explore a better fidelity.

\subsubsection{3-qubit quantum systems}

For a 3-qubit system, its Hamiltonian is assumed to be $H(t)=H_0+\sum_{m=1}^{6}u_m(t)H_m$. Denote $\sigma_x^{(12)}=\sigma_x\otimes
\sigma_x \otimes I_2 $, $\sigma_x^{(23)}= I_2\otimes \sigma_x \otimes \sigma_x$,
$\sigma_x^{(13)}= \sigma_x \otimes I_2 \otimes \sigma_x $.  The free Hamiltonian is $
H_0=0.1\sigma_x^{(12)}+0.1\sigma_x^{(23)}+0.1\sigma_x^{(13)}$.
The control Hamiltonian operators are $H_1=\sigma_x\otimes I_2 \otimes I_2 $, $H_2= I_2
\otimes \sigma_x \otimes I_2$, $H_3= I_2 \otimes I_2 \otimes \sigma_x$,
and $H_4=\sigma_z\otimes I_2 \otimes I_2 $, $H_5= I_2 \otimes
\sigma_z \otimes I_2$, $H_6= I_2 \otimes I_2 \otimes \sigma_z$.
The maximum control step is set as $N_{\max}=100$ and the control bounds are set as $G_1=G_2=G_3=G_4=G_5=G_6=1$.


\begin{figure}[ht]
	\centering
	\includegraphics[width=3.5in]{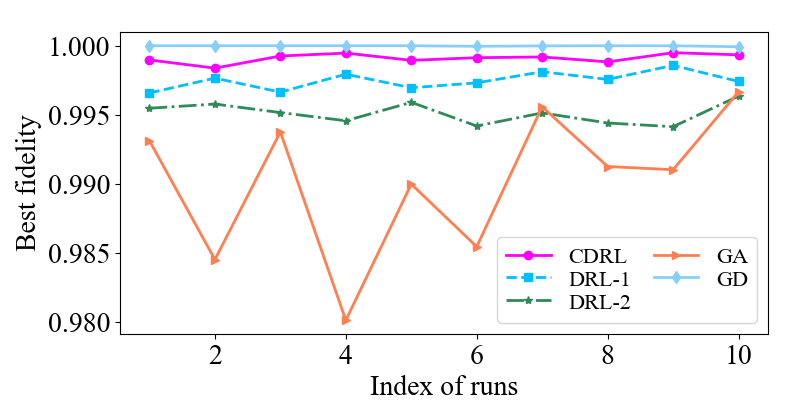}
	\caption{Comparsion of 3-qubit systems using different methods for benchmark case. }
	\label{fig:level8bench}
\end{figure}

\begin{figure}
\centering
\begin{minipage}{0.45\linewidth}
  \centerline{\includegraphics[width=3.5in]{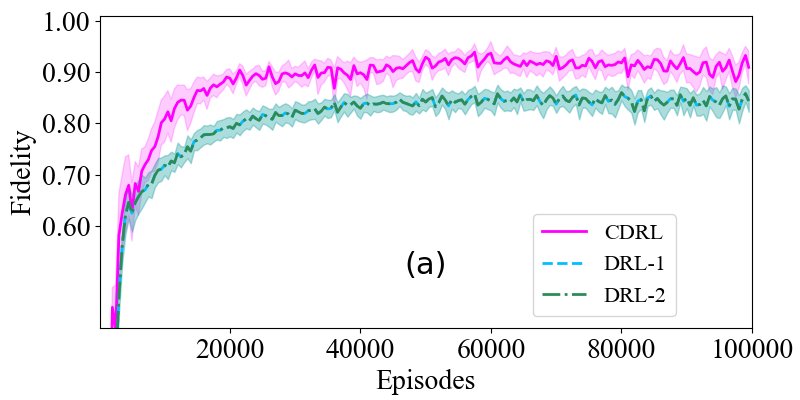}}
\end{minipage}
\vfill
\begin{minipage}{0.45\linewidth}
  \centerline{\includegraphics[width=3.5in]{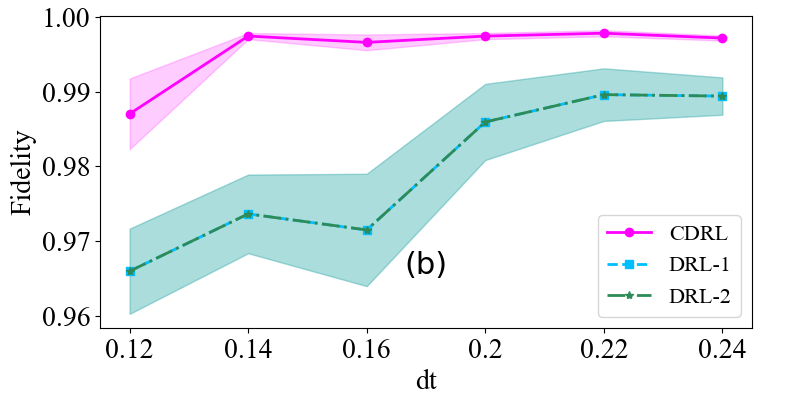}}
\end{minipage}
\caption{Comparison between CDRL and other two DRL methods, where DRL-1 and DRL-2 show the same performance in this case. (a) Fidelity of CDRL and DRL, (b) Best fidelity under different time durations.}
\label{fig:level8random}
\end{figure}


For the benchmark scenario, we take a static task-generation scheme. The numerical results under $dt=0.14$ are shown as in Fig. \ref{fig:level8bench}, revealing that CDRL achieves a comparative performance to GD and a superior result than GA. Among the three DRL methods, CDRL achieves the best performance, with DRL-1 falling far behind. 

For the random scenario, a dynamic task-generation scheme is adopted since a static one might be inappropriate for different initial states. The comparison performance between CDRL, DRL-1, and DRL-2 is summarized as in Fig. \ref{fig:level8random}. As we can see, DRL-2 with fixed target fidelity fails to achieve fidelity larger than $0.99$ and is reduced to DRL-1, and hence they show the same performance. From Fig. \ref{fig:level8random}(a), there is a wide gap between the fidelity of CDRL and that of the other two DRL methods. A further comparison regarding different time durations is summarized as in Fig. \ref{fig:level8random}(b), demonstrating that CDRL achieves the best results with different time durations.

\subsection{Numerical results for open quantum systems}\label{Sec:open}

For open quantum systems with a density operator $\rho(t)$, a real vector can be used to represent the current state in DRL and is defined as
\begin{equation}
s_j=(y_1(t),...,y_N(t))^\top, \quad y_i(t)={\rm{tr}}(Y_i \rho(t)).
\end{equation}
 Considering the complex dynamics of open systems, a dynamical task-generation mechanism is adopted in this section.

\begin{figure}[htp]
\centering
\begin{minipage}{0.4\linewidth}
  \centerline{\includegraphics[width=3.5in]{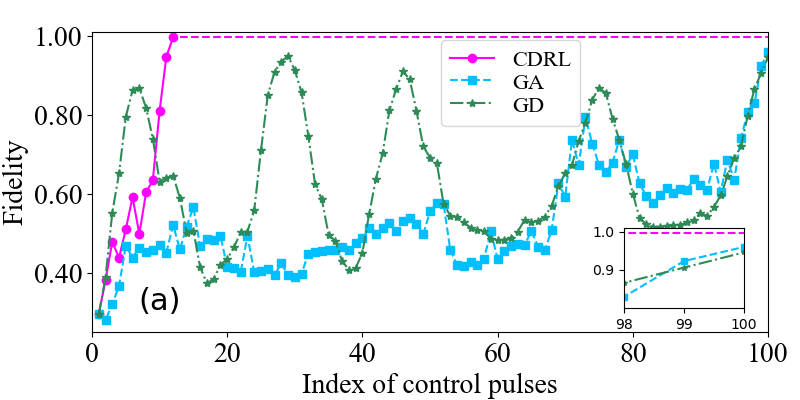}}
\end{minipage}
\vfill
\begin{minipage}{0.4\linewidth}
  \centerline{\includegraphics[width=3.5in]{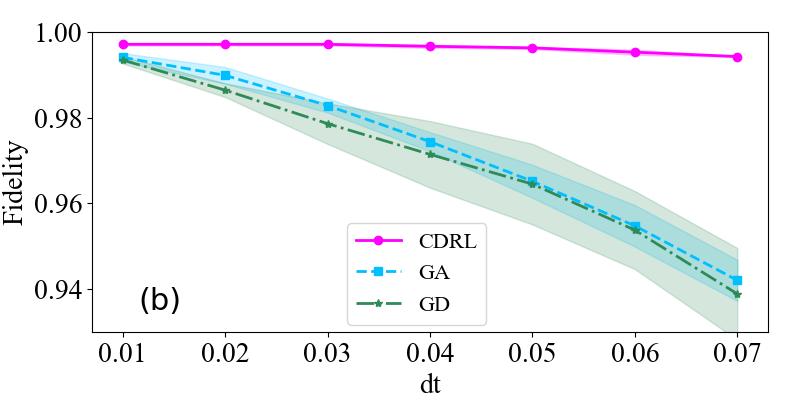}}
\end{minipage}
\caption{Comparison of CDRL, GA and GD on three-level open quantum systems.
(a) Trajectories with steps by the optimal control pulses, (b) Best fidelity under different time durations.}
\label{fig:open3EAs}
\end{figure}

\begin{figure*}[ht]
\centering
\begin{minipage}{0.25\linewidth}
  \centerline{\includegraphics[width=2.5in]{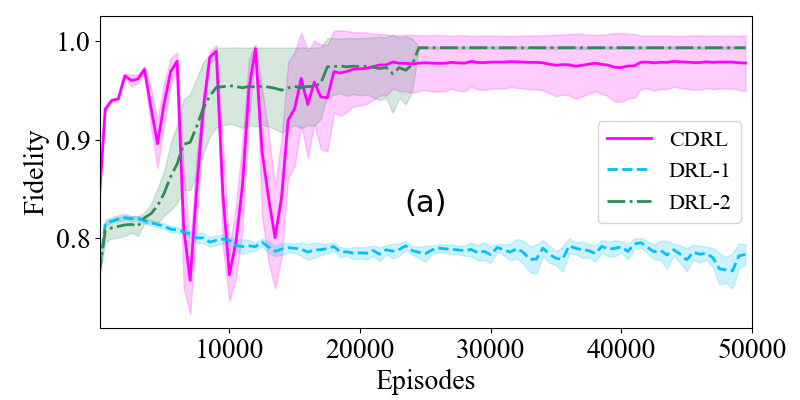}}
\end{minipage}
\hspace{1.5cm}
\begin{minipage}{0.25\linewidth}
  \centerline{\includegraphics[width=2.5in]{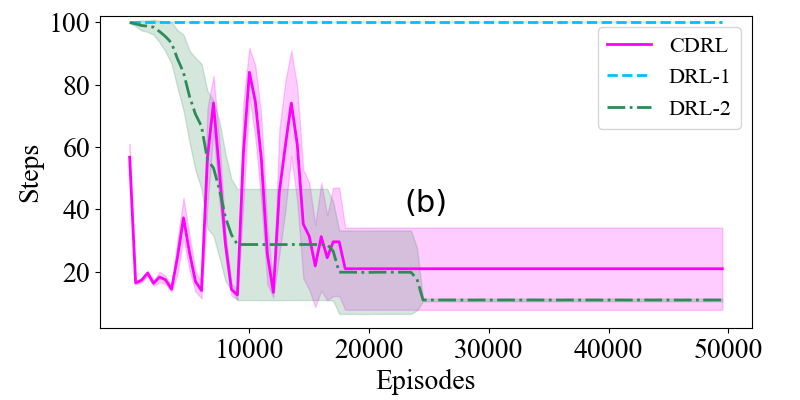}}
\end{minipage}
\hspace{1.5cm}
\begin{minipage}{0.25\linewidth}
  \centerline{\includegraphics[width=2.5in]{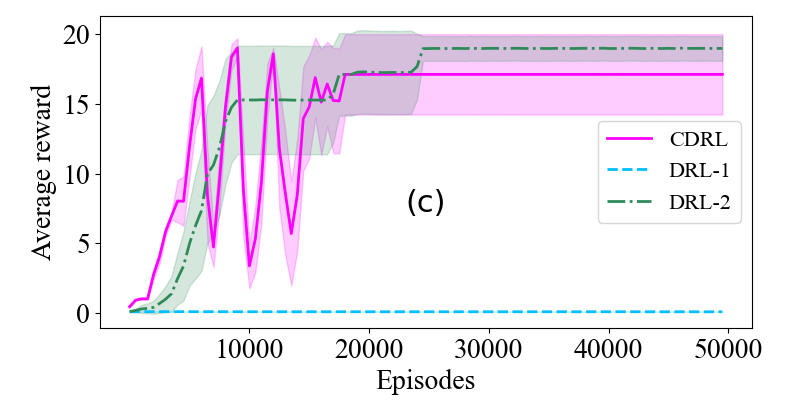}}
\end{minipage}
\caption{Training performance of three-level open quantum systems. (a) Fidelity at the end of one episode, (b) Steps in one episode, (c) Average reward for one episode.}
\label{fig:open3DRL}
\end{figure*}

\begin{figure*}[ht]
\centering
\begin{minipage}{0.25\linewidth}
  \centerline{\includegraphics[width=2.5in]{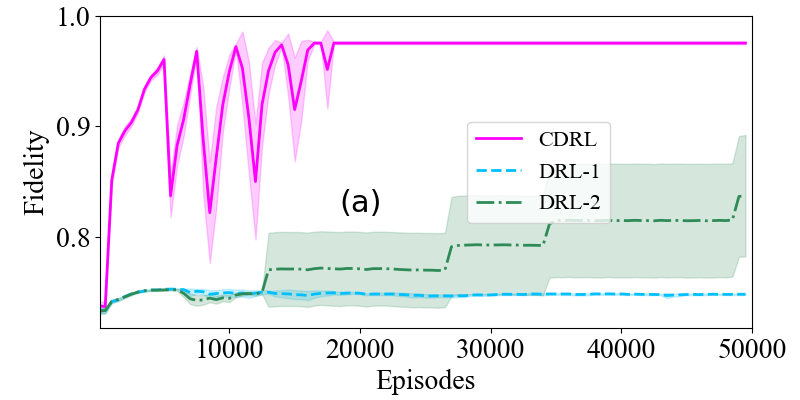}}
\end{minipage}
\hspace{1.5cm}
\begin{minipage}{0.25\linewidth}
  \centerline{\includegraphics[width=2.5in]{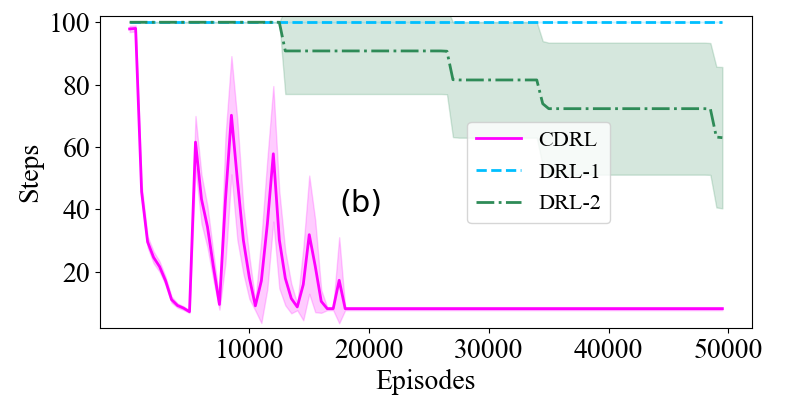}}
\end{minipage}
\hspace{1.5cm}
\begin{minipage}{0.25\linewidth}
  \centerline{\includegraphics[width=2.5in]{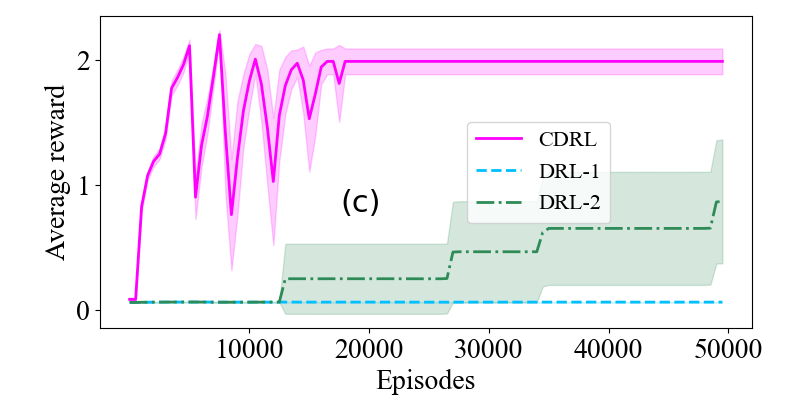}}
\end{minipage}
\caption{Training performance of four-level open systems. (a) Fidelity at the end of one episode, (b) Steps in one episode, (c) Average reward for one episode.}
\label{fig:open4DRL}
\end{figure*}

\begin{figure}[ht]
\centering
\includegraphics[width=3.5in]{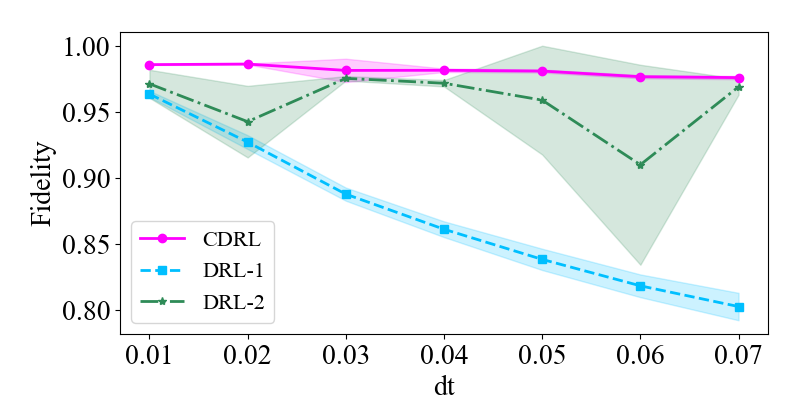}
\caption{Best fidelity under different time durations.}
\label{fig:open4time}
\end{figure}

\subsubsection{Three-level open quantum systems}

For a three-level open quantum system, let the three energy levels be $|1\rangle=(1,0,0)^\top$, $|2\rangle=(0,1,0)^\top$, $|3\rangle=(0,0,1)^\top$. The free Hamiltonian is assumed to be $H_0=\sum_{j=1}^{3}0.2 j(j+1)|j\rangle\langle j|$ and  the control Hamiltonian operators are chosen as
$H_1=\sum_{i=1}^2(|i \rangle \langle i+1|+|i+1 \rangle \langle i|)$ and $H_2=|1 \rangle \langle 3|+|3 \rangle \langle 1|$. The Lindblad operators are given as \cite{Jirari2005Optimal}		
\begin{equation}
 L_1=\tau_{12} |1\rangle \langle 2|,\quad
L_2=\tau_{13} |1\rangle \langle 3|,\quad
L_3=\tau_{23} |2\rangle \langle 3|,
\end{equation}
with $\tau_{12}=0.4$, $\tau_{23}=0.3$, $\tau_{13}=0.2$. Consider a control task from an initial state $|\psi_0 \rangle =|1\rangle$ to the target state $|\psi_f \rangle =|3\rangle$, we define a maximum step $N_{\max}=100$. 

Fig. \ref{fig:open3EAs}(a) reveals three trajectories with the optimal control pulses under $dt=0.05$. A closer look at the inset of Fig. \ref{fig:open3EAs}(a), CDRL achieves the best fidelity, and GA achieves a slightly better result than GD. The curves in Fig. \ref{fig:open3EAs}(b) demonstrate that CDRL obtains better results than GA and GD under different time durations. In addition, the curve of CDRL is close to a horizontal straight line, while GA and GD exhibit similar performance, with  fidelity decreasing with time duration. The comparison of training performance using the three DRL variants is summarized as in Fig. \ref{fig:open3DRL}, where CDRL exhibits a better performance than DRL-2 regarding fidelity. Since CDRL might explore a few steps to achieve a tiny increment of fidelity, the average steps of CDRL are  a little higher than DRL-2. While DRL-1 falls far away from CDRL and DRL-2 regarding fidelity, steps, and average reward.


\subsubsection{Four-level open quantum systems}
For a four-level open quantum system, we assume that the four energy levels are $|1\rangle=(1,0,0,0)^\top$,
$|2\rangle=(0,1,0,0)^\top$, $|3\rangle=(0,0,1,0)^\top$, $|4\rangle=(0,0,0,1)^\top$.
Let the free Hamiltonian be $H_0=0.25\sum_{j=1}^{4}j(j+1)|j\rangle\langle j|$
and the control Hamiltonian operators be $H_1=\sum_{i=1}^3(|i \rangle \langle i+1|+|i+1 \rangle \langle i|)$, $H_2=\sum_{i=1}^2(|i \rangle \langle i+2|+|i+2 \rangle \langle i|)$ and $H_3=|1 \rangle \langle 4|+|4 \rangle \langle 1|$. The Lindblad operators are given as \cite{Jirari2005Optimal}	
\begin{align}
     	&L_1=\tau_{12} |1\rangle \langle 2|,\quad
      	L_2=\tau_{13} |1\rangle \langle 3|,\quad
      	L_3=\tau_{14} |1\rangle \langle 4|,\quad
      	\\& \nonumber
      	L_4=\tau_{12} |2\rangle \langle 3|,\quad
      	L_5=\tau_{13} |2\rangle \langle 4|,\quad
      	L_6=\tau_{12} |3\rangle \langle 4|,
     	\end{align}
with $\tau_{12}=0.4$, $\tau_{13}=0.3$, $\tau_{14}=0.2$. Consider a control task from an initial state $|\psi_0 \rangle =|1\rangle$ to the target state $|\psi_f \rangle =|4\rangle$, we set the  maximum step as $N_{\max}=100$. 

The comparison result of the three DRL methods under $dt=0.06$ is summarized as in Fig. \ref{fig:open4DRL}, where CDRL converges to a better fidelity compared with the other two DRL variants in Fig. \ref{fig:open4DRL}(a). In addition, CDRL achieves the best performance regarding both steps in Fig. \ref{fig:open4DRL}(b) and average reward in Fig. \ref{fig:open4DRL}(c). The steps in an episode mean the number of control pulses for a control strategy,  suggesting that CDRL has the potential to search for control fields that can achieve higher fidelity within shorter control pulses for four-level open quantum systems compared to DRL-1 and DRL-2.  

We also implement simulations with different time durations. The comparison results are summarized as in Fig. \ref{fig:open4time}, where the performance of CDRL is much higher than that of DRL-1. In addition, the curve of CDRL is close to a horizontal straight line, while the curves of DRL-1 and DRL-2 decrease greatly with the increase of time duration.


\section{Conclusions}\label{Sec:conclusion}
In this paper, a curriculum-based deep reinforcement learning (CDRL) approach is presented
for the learning control design of quantum systems. In CDRL, a curriculum is constructed using a set of intermediate tasks. In particular, fidelity is utilized to indicate the difficulty of one task. By sequencing a set of tasks, the agent learns to grasp easy knowledge and gradually increase the difficulty until the final task is achieved. The numerical results show that CDRL achieves comparative performance to GD, and achieves better performance than GA. Moreover, CDRL exhibits superior performance
over two traditional DRL methods regarding fidelity and tends to find shorter control pulses. For our future work, we will extend this work to a sample-level curriculum to achieve more accurate manipulation of quantum systems. In addition, we will focus on the investigation of curriculum-based learning approaches for continuous control fields and integrate curriculum learning into other DRL methods such as DDPG \cite{lillicrap2016continuous} and TRPO \cite{schulman2015trust}.


\appendix
\section{Appendices}
\subsection{Pseudo code and parameter setting for GD and GA}\label{app:GDsetting}

Let the control fields be a real vector $\theta$ and denote $J(\theta)$ as the cost function for optimizing quantum systems. Considering the physical restriction of control fields, the parameters are usually initialized as $\theta=-G_i+2*\textup{Rand}[0,1]*G_i$, where $\textup{Rand}[0,1]$ is a uniform function. The algorithm description of GD is summarized as in Algorithm \ref{al:GD}. When applying GD to prepare a state on closed quantum systems, the learning rate is set as $\alpha=0.0001$. For open quantum systems, the learning rate is initialized as $\alpha=0.05$ and is decayed with $\alpha \leftarrow 0.9\alpha$ every 1000 iterations to avoid oscillations.

\begin{algorithm}[!ht]
\caption{Algorithm description for GD}\label{al:GD}
\KwIn{Learning rate $\alpha$}
\KwOut{Optimal control fields $\theta^{*}$}
Randomly initialize $\theta$\;
  \While{ not convergent}
  {
   Compute the gradient information $\frac{\partial J(\theta)}{\partial{\theta}}$\;
   Update the parameter $\theta \leftarrow \theta + \alpha \frac{\partial J(\theta)}{\partial{\theta}}$\;
  }
  Optima control fields $\theta^{*}=\theta$\;
\end{algorithm}

\begin{algorithm}[!ht]
	\caption{Algorithm description for GA}\label{al:GA}
	\KwIn{ Population size $NP$, crossing-over rate $P_c$, mutation rate $P_m$}
	\KwOut{Optimal control pulses $\theta^{*}$}
	Randomly generate $NP$ binary string $\{\vartheta_i\}$ and constitute $S=\{\vartheta_1,...,\vartheta_{NP}\}$\;
	\While{not convergent}
	{
		Map $\vartheta_i\in \{S\}$ to control fields $\theta_i=g(\vartheta_i)$\;
		Obtain fitness function $J_i=J(\theta_i)$ and rank $\vartheta_i\in \{S\}$ according to $\{J_i\}$ (descending)\;
		Select top $\lceil  NP(1-P_c)\rceil$ vectors to constitute $S_{1}$\;
		Sample $NP-\lceil  NP(1-P_c)\rceil$ vectors from $S$ with
		probability $P(\theta_i)=\frac{J_i}{\sum^{NP}_{j=1}J_j}$, to constitute $S_{2}$\;
		Randomly pair vectors among $S_{2}$ and perform $\lceil N_2/2 \rceil$ times of crossover to renew vectors in $S_{2}$\;
		Mutate vectors in $S_{2}$ with probability $P_m$\;
		Obtain new generation $S \leftarrow \{S_{1},S_{2}\}$\;  
	}
	$\theta^*= g[{\arg\max}_{\vartheta_i} J(g(\vartheta_i))]$\;
\end{algorithm}

For GA, vectors are encoded as a binary string, where $1$ corresponds to the positive control with strength $G_i$, while $0$ corresponds to the negative control with strength $-G_i$ for the $i$-th control $u_i(t)$. For convenience, a function $g(\cdot)$ is defined to map a binary string into the real control field. The algorithm description of GA is summarized as in Algorithm \ref{al:GA}, where $\lceil x \rceil$ takes the smallest integer that is not smaller than $x$. When applying GA to closed quantum systems, the populations size is set as $NP=10$, the  crossover rate and the mutation rate are set as $P_c=0.9$ and $P_m=0.0001$, respectively. For open quantum systems, the parameter settings are $NP=10$, $P_c=0.8$, and $P_m=0.005$.

\bibliographystyle{ieeetr}
\bibliography{mybib}

\end{document}